# Improved Method for Individualization of Head-Related Transfer Functions on Horizontal Plane Using Reduced Number of Anthropometric Measurements

Hugeng, Wahidin Wahab, and Dadang Gunawan

**Abstract**— An important problem to be solved in modeling head-related impulse responses (HRIRs) is how to individualize HRIRs so that they are suitable for a listener. We modeled the entire magnitude head-related transfer functions (HRTFs), in frequency domain, for sound sources on horizontal plane of 37 subjects using principal components analysis (PCA). The individual magnitude HRTFs could be modeled adequately well by a linear combination of only ten orthonormal basis functions. The goal of this research was to establish multiple linear regression (MLR) between weights of basis functions obtained from PCA and fewer anthropometric measurements in order to individualize a given listener's HRTFs with his or her own anthropometry. We proposed here an improved individualization method based on MLR of weights of basis functions by utilizing 8 chosen out of 27 anthropometric measurements. Our objective experiments' results show a superior performance than that of our previous work on individualizing minimum phase HRIRs and also better than similar research. The proposed individualization method shows that the individualized magnitude HRTFs could approximated well the the original ones with small error. Moving sound employing the reconstructed HRIRs could be perceived as if it was moving around the horizontal plane.

——————————— ◆ ———————————

## 1 INTRODUCTION

Without two eyes, direction of sound source can be recognized by a person by utilizing his or her two ears. The primary cues in localizing the direction of a sound are interaural time difference (ITD), interaural level difference (ILD), and spectral modification caused by pinna, head, and torso. These primary sound cues are encrypted in HRTF. On the horizontal plane, ITD and ILD are two main cues due to the perception of sound direction[1]. HRTF is defined as the acoustic filter of human auditory system, in frequency domain, from a sound source to the entrance of ear canal. The counterpart of HRTF in time domain is known as head-related impulse response (HRIR). One key implementation of binaural HRTFs is in the creation of Virtual Auditory Display (VAD) in virtual reality to filter monaural sound. This fact is based on the human psychoacoustic characteristic, i.e. a convincing spatial sound can be obtained sufficiently using two channels. As suggested by [3] and [4], HRTF changes with directions of sound sources and varies from subject to subject due to inter-individual difference in anthropometric measurements.

Synthesis of ideal VAD systems needs a series of empirical measurements of individual HRTFs for every listener. These measurements are not practical because of the requirements of heavy and expensive equipments as well as a long measurement time. Most commercial virtual auditory systems are recently synthesized using generic/non-individualized HRTFs that ignore inter-subject difference. However, non-individualized HRTFs suffer from distortions such as in-head localization when using headphones, inaccurate lateralization, poor vertical effects, and weak front-back distinction caused by unsuitable HRTFs applied to a listener [1],[4]. Thus, it is needed and a priority to develop an individualization method to estimate proper HRIRs for a listener, that present adequate sound cues without measurement of the individual HRIRs.

The individualization of HRTF in frequency domain or HRIR in time domain is nowadays a challenging subject of much research. Several HRTF individualization methods have been developed, such as HRTF clustering and selection of a few most representative ones [5], HRTF scaling in frequency [6], a structural model of composition and decomposition of HRTFs [7], HRTF database matching [8], the boundary element method [9], HRIR subjective customization of pinna responses [10] and of pinna, head, and torso responses [11] in the median plane, and HRTF personalization based on multiple regression analysis (MRA) in the horizontal plane [12]. Shin and Park [10] suggested HRIR customization method based on subjective tuning of

————————————————
- *Hugeng is with the Department of Electrical Engineering, University of Indonesia, Depok 16424 – Indonesia.*
- *W. Wahab is with the Department of Electrical Engineering, University of Indonesia, Depok 16424 – Indonesia.*
- *D. Gunawan is with the Department of Electrical Engineering, University of Indonesia, Depok 16424 – Indonesia.*

only pinna responses (0.2 ms out of entire HRIR) in the median plane using PCA of the CIPIC HRTF Database [2]. They achieved the customized pinna responses by letting a subject tune the weight on each basis function. Hwang and Park [11] follow the similar method as [10], but they fed PCA with the entire median HRIRs; each HRIR is 1.5 ms long (67 samples) since the arrival of direct pulse. This HRIR includes the pinna, head, and torso responses. They tuned subjectively the weights of three dominant basis functions due to the three largest standard deviations at each elevation. Hu et al. [12] personalized the estimated log-magnitude responses of HRTFs by MRA. At the beginning, the log-magnitude responses are estimated using PCA as linear combination of weighted basis functions. The weights of the basis functions are then estimated using anthropometric measurements based on MRA.

Our individualization method is similar to the method in [12], but we employed in the PCA modeling, the magnitude responses of HRTFs, instead of the log-magnitude responses of HRTFs utilized by Hu et al., however, our selection procedure of anthropometric measurements is also different. Entire horizontal magnitude HRTFs calculated from the original HRIRs in the CIPIC HRTF Database are included in a single analysis. Thus, all horizontal magnitude HRTFs for both ears share the same set of basis functions, which cover not only the inter-individual variation but also the inter-azimuth variation. This paper presents an individualization method by developing the statistical PCA model of magnitude HRTFs and MLR between weights of basis functions and selected few anthropometric measurements, that was different and showed improved performance from [12].

Section 2 describes the proposed algorithm of individualization method, database used, minimum phase analysis, PCA of magnitude HRTFs, minimum phase reconstruction and synthesis of HRIR models, individualization of magnitude HRTFs using MLR, and correlation analyses for the selection process of independent variables and dependent variables of MLR models. Section 3 discusses experiments' results, which consist of discussions of resulted basis functions and weights of basis functions from PCA, and the performance of the proposed individualization method.

## 2 PROPOSED INDIVIDUALIZATION METHOD

The goal of our research is to develop an improved individualization method of HRTFs on the horizontal plane, by using multiple regression models between magnitude HRTFs and a few anthropometric measurements. This method individualizes magnitude HRTF models into suitable HRIRs for a given listener, by using a few of his or her own anthropometric measurements. The suitable individualized HRIRs are necessary when the listener uses a spatial audio application.

The schematic diagram of the proposed HRTFs individualization method is shown in Fig. 1. The database of HRIRs used in the research was provided by CIPIC Interface Laboratory of California University at Davis [2], [3].

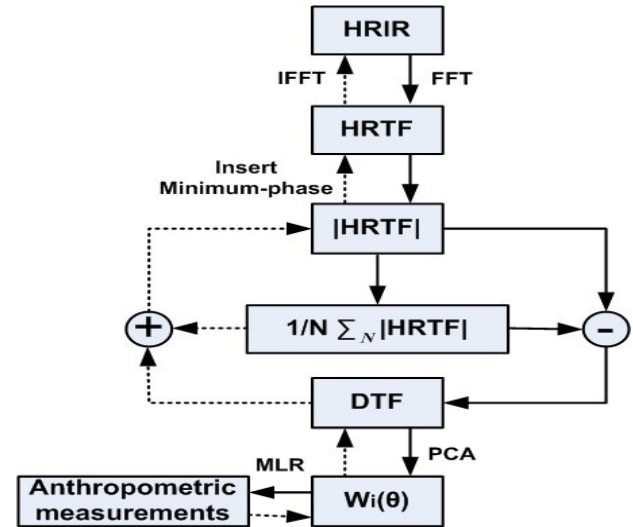

Fig. 1. Proposed HRTFs Individualization Method.

This database is reviewed briefly in the subsection below. Firstly, as seen in Fig. 1, we obtained from the database the entire original HRIRs on horizontal plane of 37 subjects, which consists of 50 HRIRs of each ear and each subject. We used a total number of 3700 HRIRs in modeling and individualizing HRIRs of a listener. Each HRIR was processed by 256-points fast Fourier transform (FFT) to transform it into its corresponding complex HRTF.

As the object of HRTF modeling using PCA, we took only 128 frequency components of magnitude of the complex HRTF. At this step, the phase of the complex HRTF was discarded. Then, we computed the mean of the entire magnitude HRTFs. This mean was substracted from each magnitude HRTF to obtain its corresponding direct transfer function (DTF). This substraction was performed in order to have centered data of magnitude HRTF, called DTF, which was necessary for PCA to get a good result.

For HRTFs modeling purpose, all DTFs were thus fed into PCA. The PCA delivered 128 ordered basis functions or principal components (PCs) and their weights (PCWs). The PCs were ordered from the PC with largest eigen value to the PC with smallest eigen value. It must be kept in mind that each eigen value determined the percentage variance of all DTFs explained by its corresponding PC. The first PC that corresponds to the largest eigen value explained largest percentage variance of the entire DTFs. To attain later individualized HRTFs of a new listener, we performed multiple linear regression (MLR) between the PCWs resulted from PCA and a few anthropometric measurements of 37 subjects in the database. Detailed selection process of anthropometric measurements from a total of 27 measurements is explained in the separated subsection below. The MLR method provided regression coefficients that correlated the PCWs and selected anthropometric measurements. These regression coefficients were thus applied to a set of anthropometric measurements of a new listener to obtain estimated PCWs for that listener. A linear combination of weighted PCs using these estimated PCWs resulted in an individualized DTF.

The desired individualized HRIRs of a listener were attained using the reconstruction process shown by the dashed lines in Fig. 1. Each individualized DTF that was achieved from the MLR method and PCA, was added to the mean of DTFs calculated before to yield its individualized magnitude HRTF. Minimum-phase was inserted to the individualized magnitude HRTF to result in an individualized complex HRTF. Here we followed the assumption that the phase of the HRTF can be approximated by minimum-phase [13]. The inverse Fourier transform finally was applied to obtain individualized HRIRs from the corresponding complex HRTFs. The initial left- and right-ear time delay due to the distance from the sound source in a particular direction to each ear drum were inserted respectively to the left-ear HRIR and to the right-ear HRIR.

The database used, the minimum phase analysis, reconstruction, PCA of the magnitude HRTFs in the frequency domain, minimum phase reconstruction and synthesis of HRIRs, MLR method, and selection of anthropometric measurements are explained in the following subsections.

## 2.1 The Database Used

Most commercial VAD systems convolve input signals with a pair of standard HRIRs, which ordinarily come from a serial of studies that used public HRIR data of acoustic manikin called Knowles Electronics Manikin for Auditory Research (KEMAR). HRIRs vary significantly among individuals, hence a database which results from sufficiently large number of HRIRs measurements is needed in order to perform HRIRs modeling. CIPIC Interface Laboratory at California University, Davis - USA, had measured HRIRs with a high spatial resolution from more than 90 subjects [2],[3]. They has released CIPIC HRTF Database Release 1.2, which is a subset of database for only 45 subjects. This database is downloadable from their website and can be used freely for academic research purpose. CIPIC HRTF Database not only consists of impulse responses from 1250 spatial directions for each ear and each subject, but also includes a set of anthropometric measurements of all subjects. We used the CIPIC HRTF Database in our research because of its extent features.

The number of subjects involved in the HRIRs measurements is 43 people, consists of 27 males and 16 females. Two other subjects are KEMAR with small pinnae and KEMAR with large pinnae. All impulse responses were measured with condition that the subject sat in the center of a circle with radius 1 meter. The position of head was not fixed, but the subject could monitor his or her head position.

As sound sources, Bose Acoustimass™ loudspeakers, with cone diameter of 5.8 cm, were mounted at different positions on the half-circled hoop. Golay-code signals were generated by a modified Snapshot™ system from Crystal River Engineering. Each ear canal was blocked and '*Etymotic Research ER-7C*' probe microphones were used to pick up the Golay-code signals. Output of a microphone was sampled with frequency 44,100 Hz, 16 bit resolution, and processed by Snapshot's oneshot function to produce a raw HRIR. A modified Hanning window was applied on the raw HRIR to eliminate room reflections and then the result was free-field compensated to improve the spectral charateristics of the transducers used. The length of each HRIR is 200 samples with duration of about 4.5 ms.

Direction of a sound source was determined by azimuth angle, $\theta$, and elevation angle, ø, in interaural-polar coordinate system. Elevation was sampled at $360/64 = 5,625°$ step from -45° to +230.625°, while azimuth was sampled at -80°, -65°, -55°, from -45° to +45° in 5° step, at 55°, 65°, and 80°. Hence, 1250-points spatial samples were obtained from the measurements of each ear of a subject. The published CIPIC HRTF Database contains anthropometric measurements of each subject. Although these measurements are not very accurate, but they allow investigation about possible correspondence or correlation among physical dimensions and HRTF characteristics. Following the approach suggested by Genuit [5], there are 27 anthropometric measurements in

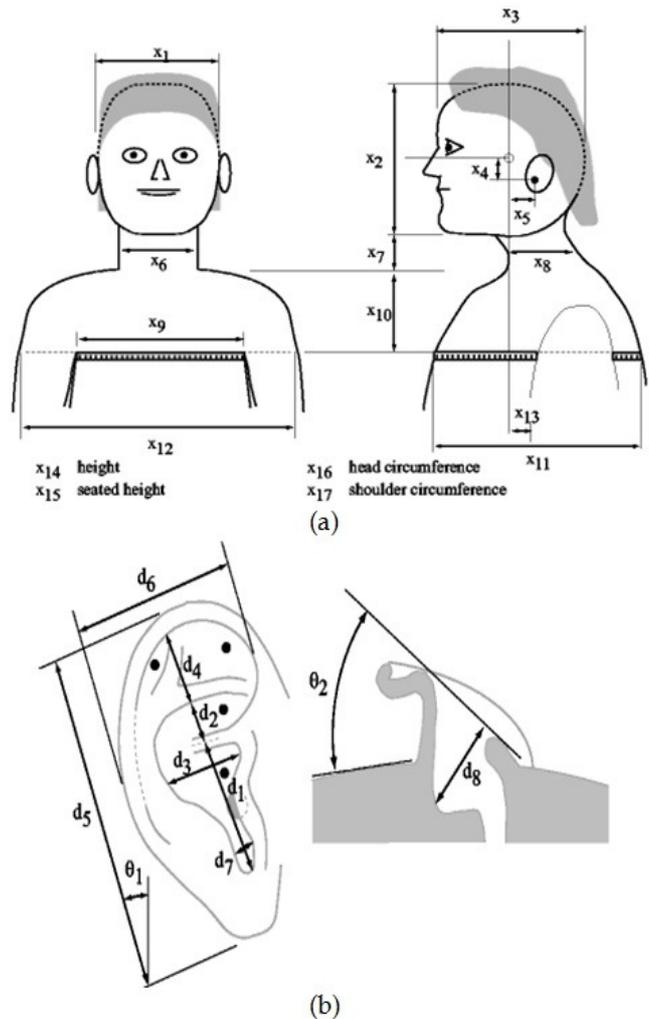

Fig. 2. Subject's Anthropometric Measurements:

(a) of Head and Torso, (b) of Pinna [2],[3].

the database, which consists of 17 measurements of head and torso, and 10 measurements of pinna as shown in Fig. 2 [2], [3]. Generally, histogram of subjects' measurements indicates a normal distribution of values. Discarding the offset measurements $x_4$, $x_5$, $x_{13}$, where the percentages of deviation can be ignored, mean of percentages of deviation is ±26%. Thus, there are a sufficient number of variations in the measurements and sizes of subjects in the database used.

## 2.2 Minimum Phase Analysis

Each HRIR in the dababase used was measured with a distance of one meter from the sound source to the center of subject's head. From the graph of HRIR versus time, it is observed a time delay due to the distance mentioned before, which is needed by sound wave to propagate from its source to the ear drum, before a maximum amplitude of HRIR occurs. To eliminate this time delay, HRIR can be reconstructed into a minimum-phase HRIR using Hilbert transform. In the minimum-phase HRIR, the phase is allowed to be arbitrary or else it is set in such a way that the magnitude response of HRIR is made easier to achieve. A linear time invariant filter, $H(z) = B(z)/A(z)$, is said to have minimum phase if all of its poles and zeros are inside the unit circle, $|z|=1$, in the z-plane. Equivalently, a filter $H(z)$ has minimum phase if not only itself but also its inverse, $1/H(z)$, are stable. A minimum phase filter is also causal since noncausal terms in the transfer function correspond to poles at infinity. The simplest example of minimum phase filter would be the unit-sample advance, $H(z) = z$, which consists of a zero at $z = 0$ and a pole at $z = \infty$. A filter is called to have minimum phase if both the numerator and denominator of its transfer function are minimum phase polynomials in $z^{-1}$, i.e. a polynomial of the form,

$$B(z) = b_0 + b_1 z^{-1} + b_2 z^{-2} + \ldots + b_M z^{-M}$$
$$= b_0(1-\beta_1 z^{-1})(1-\beta_2 z^{-1})\ldots(1-\beta_M z^{-1}) \quad (1)$$

is said to have minimum phase if all of its roots, $\beta_i$, i=1,2,...,M, lie inside the unit circle, i.e. $|\beta_i|<1$. A general property of minimum phase impulse responses is that among all impulse responses, $h_i(n)$, having identical magnitude spectra, impulse responses with minimum phases experience the fastest decay in the sense that,

$$\sum_{n=0}^{K}|h_{mp}(n)|^2 \geq \sum_{n=0}^{K}|h_i(n)|^2, \; n=0, 1, 2, \ldots, K, \quad (2)$$

where $h_{mp}(n)$ is a minimum phase impulse response.

The equation above represents that the energy in the first K + 1 samples of the minimum-phase case is at least as large as any other causal impulse response having the same magnitude spectrum. Thus, minimum-phase impulse responses are maximally concentrated toward time t=0 among the space of causal impulse responses for a given magnitude spectrum. Because of this property, minimum-phase impulse responses are sometimes called minimum-delay impulse responses. It is known that in a minimum phase filter, $H(z) = e^{a(z)} e^{i\,b(z)}$, the relations, $b(z) = -\mathcal{H}\{a(z)\}$ and $a(z) = -\mathcal{H}\{b(z)\}$, are also valid, where $\mathcal{H}\{\}$ is the Hilbert transform. The logarithmic change of these relations was obtained mainly through the calcultion of real cepstrum.

It is proposed by Kulkarni et al. [13], that the phase of HRIR can be approximated by minimum phase. A minimum phase system function, H(z), of an HRIR, h(n), has all poles and all zeros that are placed inside the unit circle $|z|=1$ in the z-plane. The calculation of real cepstrum of an original HRIR, which has arbitrary phase, results in a minimum phase HRIR, $h_{mp}(n)$. We can say that the minimum phase HRIR is the removed initial time delay version of the correspond original HRIR. But both kinds of HRIR have the same magnitude spectrum in the frequency domain. The real cepstrum, v(n), of HRIR, h(n), is calculated as follow,

$$v(n) = \mathrm{Re}\{F_D^{-1}\{\ln|F_D\{h(n)\}|\}\}, \quad (3)$$

where ln and Re{} denote respectively natural logarithm and the real part of a complex variable, $F_D\{\}$ and $F_D^{-1}\{\}$ are the discrete Fourier transform and its inverse respectively. This real cepstrum is then weighted by the following window function,

$$w(n) = \begin{cases} 0 & \text{if } n < 0, \\ 1 & \text{if } n = 0, \\ 2 & \text{if } n > 0. \end{cases} \quad (4)$$

In case of a rational H(z), the window function can be seen as a complex conjugate inversion of the zeros outside the unit circle, so that a minimum phase HRIR is provided. Hence the desired minimum phase HRIR, $h_{mp}(n)$, is resulted from:

$$h_{mp}(n) = \mathrm{Re}\{\exp(F_D\{w(n).v(n)\})\}. \quad (5)$$

## 2.3 PCA of Magnitude HRTFs in Frequency Domain

Complex HRTFs were attained by implementing fast Fourier transform (FFT) to HRIRs of the database used. The entire complex HRTFs were computed from left-ear and right-ear HRIRs of 37 subjects on horizontal plane. There are 50 HRIRs from different directions (50 azimuths) on horizontal plane for each ear of a subject, so that a total of 3700 complex HRTFs were produced by 256-points FFT. We took only magnitudes of all complex HRTFs as the input of PCA modeling. Only 128 first frequency components of a magnitude HRTF were taken into analysis because of the symmetry property of a magnitude spectrum.

A matrix composed of DTFs is needed by PCA. The original data matrix, **H** (NxM), is composed of magnitudes of HRTFs on horizontal plane, in which, each column vector, $h_i$ (i=1,2,…,M), represents a magnitude HRTF of an ear of a subject in a direction on horizontal plane. The number of magnitude HRTFs of each subject on horizontal plane is 100 (2 ears x 50 azimuths). Hence, the size of **H** is 128 x 3700 (N=128, M=3700). The empirical mean vector ($\mu$: Nx1) of all magnitude HRTFs is given by,

$$\mu = (1/M) \sum_{i=1}^{M} h_i. \qquad (6)$$

The DTFs matrix, **D**, is the mean-subtracted matrix and is given by,

$$D = H - \mu.y, \qquad (7)$$

where **y** is a 1xM row vector of all 1's. The next step is to compute a covariance matrix, **S**, that is given by

$$S = D.D^*/ (M-1) \qquad (8)$$

where * indicates the conjugate transpose operator. The basis functions or PCs, $v_i$ (i=1,2,…,q), are the q eigenvectors of the covariance matrix, **S**, corresponding to q largest eigenvalues. If q = N, then the DTFs can be fully reconstructed by a linear combination of the N PCs. However, q is set smaller than N because the goal of PCA is to reduce the dimension of dataset. An estimate of the original dataset is obtained here by only 10 PCs, which account for 93.93% variance in the original data **D**. By using only 10 PCs to model magnitude HRTFs, we expected to obtain satisfactory good results. The PCs matrix, **V** = [$v_1$ $v_2$ … $v_N$], that consisted of complete set of PCs can be obtained by solving the following eigen equation,

$$S V = \Lambda V \qquad (9)$$

where $\Lambda$ = diag{$\lambda_1$,…,$\lambda_{128}$}, is a diagonal matrix formed by 128 eigen values, where each eigen value, $\lambda_i$, represents sample variance of DTFs that was projected onto i-th eigen vektor or PC, $v_i$.

Then, the weights of PCs (PCWs), **W**(10x3700), that correspond to all DTFs, **D**, can be obtained as,

$$W = V^*.D, \qquad (10)$$

where PCs matrix now was reduced to **V** = [$v_1$ $v_2$ … $v_{10}$]. PCWs represent the contribution of each PC to a DTF. They contain both the spatial features and the inter-individual difference of DTF. Thus, the matrix consisted of models of magnitude HRTFs, $\hat{H}$, is given by,

$$\hat{H} = V.W + \mu.y. \qquad (11)$$

Tabel 1 shows the percentage variance and the cummulative percentage variance of DTFs in the database explained by PC-1 to PC-20 ($v_1$, $v_2$, … , $v_{20}$) respectively. The application of more PCs would reduce the modeling error between the magnitude HRTF of database and the model of magnitude HRTF, but on the other hand, it costed more computing time and larger memory space. The PCs-matrix, **V**, that at first has 128x128 elements was reduced into a matrix of only 128x10 elements. We used only the first 10 PCs out of all 128 PCs. In this way automatically we needed only 10 PCWs to perform the model. Hence, one can see obviously the advantage of PCA in reducing significantly the memory space needed.

TABLE 1
The Percentage of Variance Explained by Basis Functions

| PC | Eigen Value | Variance (%) | Cumulative Variance (%) |
|---|---|---|---|
| $v_1$ | 52.95 | 60.97 | 60.97 |
| $v_2$ | 9.38 | 10.80 | 71.78 |
| $v_3$ | 6.90 | 7.95 | 79.73 |
| $v_4$ | 3.38 | 3.89 | 83.62 |
| $v_5$ | 2.65 | 3.05 | 86.67 |
| $v_6$ | 1.97 | 2.27 | 88.93 |
| $v_7$ | 1.61 | 1.85 | 90.79 |
| $v_8$ | 1.07 | 1.23 | 92.02 |
| $v_9$ | 0.94 | 1.08 | 93.10 |
| $v_{10}$ | 0.72 | 0.83 | 93.93 |
| $v_{11}$ | 0.70 | 0.80 | 94.74 |
| $v_{12}$ | 0.67 | 0.77 | 95.50 |
| $v_{13}$ | 0.53 | 0.61 | 96.11 |
| $v_{14}$ | 0.36 | 0.42 | 96.53 |
| $v_{15}$ | 0.32 | 0.37 | 96.90 |
| $v_{16}$ | 0.26 | 0.30 | 97.19 |
| $v_{17}$ | 0.24 | 0.28 | 97.47 |
| $v_{18}$ | 0.22 | 0.26 | 97.72 |
| $v_{19}$ | 0.20 | 0.23 | 97.95 |
| $v_{20}$ | 0.19 | 0.22 | 98.17 |

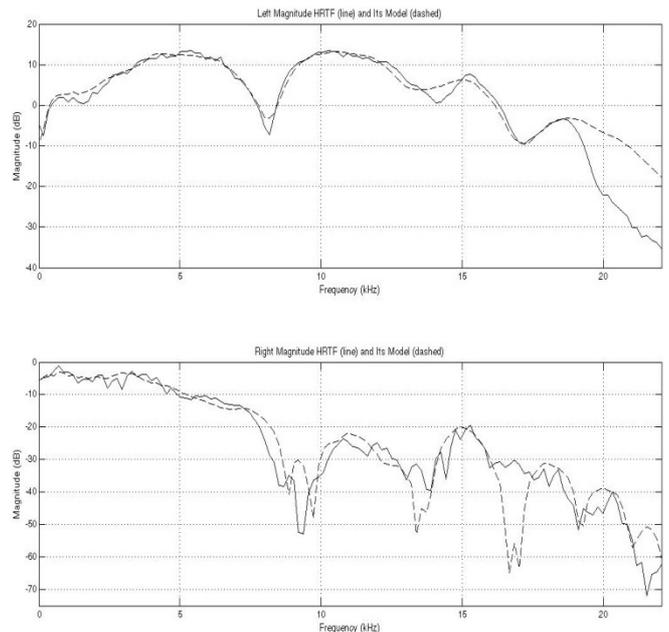

Fig. 3. Comparison of Original Magnitude HRTFs and

Their Corresponding PCA Models

Fig. 3 shows a left magnitude HRTF of Subject 003 and its PCA model due to direction with azimuth -80° and elevation 0° (top panel). On the bottom panel, it is shown the right magnitude HRTF and its PCA model due to the same direction. We can see that the models approximate well the corresponding magnitude HRTFs.

## 2.4 Minimum Phase Reconstruction and Synthesis of HRIR Models

As explained in the previous subsection, we obtained PCs matrix, **V**, and PCWs matrix, **W**, from the PCA method. Both matrices together with the empirical mean vector, **μ**, were applied to yield the matrix of models of magnitude HRTFs, $\hat{\mathbf{H}}$, as suggested by (11). By now, we could calculate the models of magnitude HRTFs of both ears. In order to synthesize the models of complex HRTFs, the phase information of left- and right-ear model of magnitude HRTF should be inserted into those models. We reconstructed the models of complex HRTFs based on the approach made by Kulkarni et al. [13]. They assumed that the phase of a HRTF was minimum phase. The phase function for a given model of magnitude HRTF was calculated by using Hilbert transform of natural logarithm of the model of magnitude HRTF. The minimum phase, $\varphi_{mp}$, of a model of magnitude HRTF, $\hat{\mathbf{h}}_i$ ((i=1,2,…,M)), is given by,

$$\varphi_{mp} = \text{Imag}\{\mathcal{H}\{-\ln(\hat{\mathbf{h}}_i)\}\}, \quad (12)$$

where Imag{} denotes the imaginary part of a complex number and ln is the natural logarithm.

Thus, the model of minimum phase complex HRTF, $\hat{\mathbf{h}}_c$, can be calculated using,

$$\hat{\mathbf{h}}_c = \hat{\mathbf{h}}_i \cdot \exp(j \cdot \varphi_{mp}), \quad (13)$$

where exp() denotes the exponential function. And the corresponding model of minimum phase HRIR, $\hat{\mathbf{h}}_{mp}(n)$, is given by the inverse fast Fourier transform (IFFT) of its complex HRTF, $\hat{\mathbf{h}}_c$, from (13). Furthermore, in reconstructing the model of left-ear minimum phase HRIR and the model of right-ear minimum phase HRIR for a particular direction of sound source into related model of left-ear HRIR and model of right-ear HRIR respectively, we needed to insert respective time delay related to the distance travelled by sound wave from the sound source to each ear drum of a subject, into each model of minimum phase HRIR. The time delays to be inserted were obtained from the means of time delays of respective directions on the horizontal plane from all subjects in the database used. The difference between left-ear time delay and right-ear time delay is called interaural time difference (ITD), which is needed by human to determine sound source direction. Fig. 4 shows, on the left panel, the original HRIRs of subject 003 due to direction with azimuth -80° and elevation 0°. On the right panel, we can see related models of left and right HRIR. These models resulted from the reconstructions of the PCA models of magnitude HRTFs into their corresponding HRIRs, as explained before. However, the models of magnitude HRTFs attained had not been individualized.

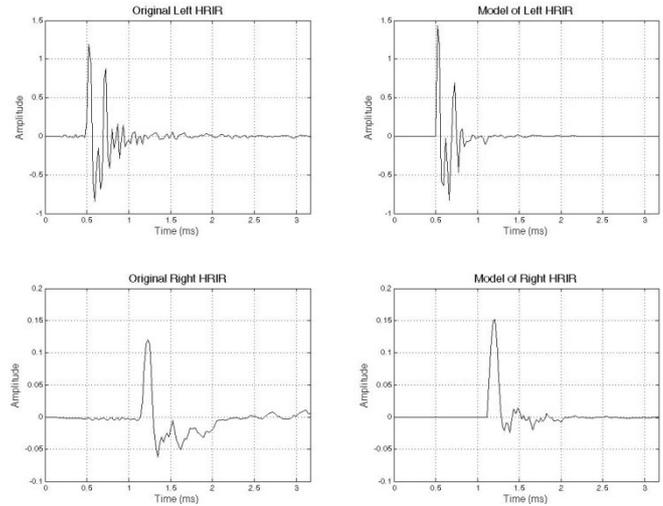

Fig. 4. Comparison of Original HRIRs and Models of HRIRs Obtained by Reconstruction of Models of Magnitude HRTFs.

## 2.5 Individualization of Magnitude HRTFs Using MLR

As shown in Fig. 1, the individualization of the models of magnitude HRTFs, which were resulted from PCA, were done through MLR of PCWs matrix, **W**, using anthropometric measurements of a listener. From the matrix **W** of (10), we can extract a weights vector, $\mathbf{w}_{i,\theta}$ (37x1), which is a vector consisted of the i-th weights of the i-th PC, $\mathbf{v}_i$, of an ear of all subjects with azimuth $\theta$ on the horizontal plane, where i=1,2,…,10. In this research, we employed only 8 anthropometric measurements of a subject in the individualization process. The selection of these 8 measurements will be discussed in detail in the separate subsection below. These selected measurements of all subjects being analyzed were then gathered together in the columns of an anthropometric matrix, **X** (37x9), where the first column of **X** consists of all 1's.

Suppose that the relation between the weights vector, $\mathbf{w}_{i,\theta}$, and the anthropometric matrix, **X**, is given by,

$$\mathbf{w}_{i,\theta} = \mathbf{X} \cdot \boldsymbol{\beta}_{i,\theta} + \mathbf{E}_{i,\theta}, \quad (14)$$

where $\boldsymbol{\beta}_{i,\theta}$ (9x1) is the regression coefficients vector and $\mathbf{E}_{i,\theta}$ (9x1) is the estimation errors vector. The regression coefficients were found by implementing least-square estimation. This estimation is performed by solving the optimization problem min{$E_{i,\theta}(n)$}, where $E_{i,\theta}(n)$ is the n-th dependent variable's estimation error. PCWs and anthropometric measurements are respectively the model's dependent and independent variables.

From (14), the regression coefficients due to i-th PCWs in azimuth $\theta$, $\mathbf{B}_{i,\theta}$, can be estimated as,

$$\mathbf{B}_{i,\theta} = (\mathbf{X}^T \cdot \mathbf{X})^{-1} \cdot \mathbf{X}^T \cdot \mathbf{w}_{i,\theta}. \quad (15)$$

As suggested by (15), enhancing the performance of the MLR method, it is needed to select both dependent and independent variables carefully. By applying PCA on magnitude HRTFs, the dimensions of independent variables were reduced significantly, so was the complexity of the models. Many correlation analyses were employed to select the independent variables in obtaining more accurate and simpler MLR method, as explained further in the subsection 2.6.

### 2.6 Correlation Analyses for Selection of Anthropometric Measurements

We employed the CIPIC HRTF Database, which are composed of both the measured HRIRs and some anthropometric measurements for 45 subjects, including the KEMAR mannequin with both small and large pinna. The detail definitions of the all 27 anthropometric measurements are given in [2], [3] and can be seen in Fig. 2. Modeling of the listener's own HRIRs via his or her own anthropometric measurements will directly affect the feasibility and complexity of the system. It is obviously not advisable to implement all measurements into the model. Some useful information will be concealed by the unnecessary measurements, which results in a worse regression model. Besides, many measurements are very difficult to be measured correctly.

There are three parameters that psychoacoustically important in the perception of natural sound, i.e. interaural time difference (ITD), interaural level difference (ILD) and pinna notch frequency, $f_{pn}$. ITD is the time difference between the arrival of first pulse of sound source from a particular direction on the left ear drum and that of the right ear drum. At the directions of sound source on median plane, ITD is near zero, where for a perfect symmetric head, there is no ITD on that plane. Thus, one can say that ITD is a function of azimuth on planes with fixed elevation. ITD can be calculated from the time delay of maximum cross correlation of the left HRIR and right HRIR at a particular direction. Then, ILD is defined as level or magnitude difference (in dB) in frequency domain between the left magnitude HRTF and the right magnitude HRTF at a particular direction of sound source. For a particular direction, we obtained ILD from each frequency component in the range of 0 – 22050 Hz. ILDs generally are analyzed for a determined frequency component on the horizontal plane and on the median plane. Another significant psychoacoustic parameter is pinna notch frequency, $f_{pn}$. Pinna notch frequency is the notch frequency in the magnitude spectrum of HRTF caused by diffraction and reflection of sound wave on a pinna.

ITD and ILD are significant for the perception of azimuth of sound source. They affect much the variation of HRTF on the horizontal plane. But ILD and $f_{pn}$ play important role in the perception of elevation of sound source and affect the variation of HRTF on the median plane. It is difficult to characterize the range of HRTF variation among subjects. However, maximum ITD, $ITD_{max}$, maximum ILD, $ILD_{max}$, and $f_{pn}$ are simple and perceptually relevant parameters that characterize existing HRTF variation. Correlation analyses were applied to determine which anthropometric measurements have strong correlations with $ITD_{max}$, $ILD_{max}$, and $f_{pn}$. From a few strongest correlated anthorpometric measurements, four measurements were chosen from head and torso sizes, i.e. $x_1$ with $\rho$ = 0.736, $x_3$ with with $\rho$ = 0.706, $x_6$ with $\rho$ = 0.726, and $x_{12}$ with $\rho$ = 0,768, where $\rho$ denotes the correlation coefficient between the measurement and $ITD_{max}$. These 4 measurements were employed in the individualization of magnitude HRTFs using MLR method. Correlation analyses between $ILD_{max}$ and head and torso sizes provided weaker correlations but confirmed the chosen of $x_1$, $x_6$, and $x_{12}$. The selection of $x_3$, $x_6$, and $x_{12}$ was also confirmed with the correlation analyses on the horizontal plane between the first PCWs, $w_{1,\theta}$, from the PCA of magnitude HRTFs and anthropometric measurements. We focused on first PCWs because they have largest variation through the azimuths.

The effects of pinna sizes are stronger with HRTFs on the median plane than HRTFs on the horizontal plane [1]. But overall the pinna sizes affect HRTFs in all directions. The correlation analyses between $f_{pn}$ and anthropometric measurements provided in general weaker correlations than those of $ITD_{max}$. Four pinna sizes had strongest correlations with $f_{pn}$ and that's why to be chosen; i.e. $d_1$ with $\rho$ = 0.435, $d_3$ with $\rho$ = 0.360, $d_5$ with $\rho$ = 0.204, and $d_6$ with $\rho$ = 0.280. These selected sizes of pinna are easy to be measured and represent measures of height and width. Hence, eight anthropometric measurements, $x_1$, $x_3$, $x_6$, $x_{12}$, $d_1$, $d_3$, $d_5$, and $d_6$ were chosen and fed in the MLR method in order to calculate regression coefficients. These eight anthropometric measurements are the same as the measurements that we used in our previous work[14]. Then the regression coefficients were applied in estimating the PCWs of a DTF at each direction on the horizontal plane.

## 3 EXPERIMENTS' RESULTS AND DISCUSSION

In this section, we discussed the performance of the proposed individualization method from the objective simulation experiments between the original magnitude HRTFs of the database and the individualized models of magnitude HRTFs. The experiments were done by employing only the data on the horizontal plane of 37 subjects out of all 45 subjects in the database. This occurred because the database had not included the complete set of anthropometric measurements of all subjects and the selected 8 anthropometric measurements were included only for 37 subjects.

### 3.1 Basis Functions Resulted from PCA

The inputs of the PCA were 3700 DTFs processed from HRIRs on the horizontal plane of 37 subjects. By solving eigen equation, we attained 10 basis functions or PCs to model the given DTFs. Fig. 5 shows the first five basis functions, $v_1,...,v_5$. As shown in this figure, that all five basis functions can be said roughly constant and approx-imate zero at frequencies below 1-2 kHz. This reflects the fact that there is almost no direction-dependent variability in the DTFs in this frequency range. Regardless of the weights employed to the basis functions, the resulting weighted sum will be close to zero in this range.

Above about 2 kHz, all five basis functions have non-zero values. It is obvious that with the exception of the first PC, the high-frequency variability in these basis functions represents the direction-dependent high-frequency peaks and notches in the DTFs. The higher order basis function has more ripples and more details especially for the frequencies above about 2 kHz. The trends explained above are similar for the sixth to tenth basis functions. Taken together, all basis functions seem to capture the high frequency spectral variability. They also reflect spectral differences between sources in front and sources behind the subject.

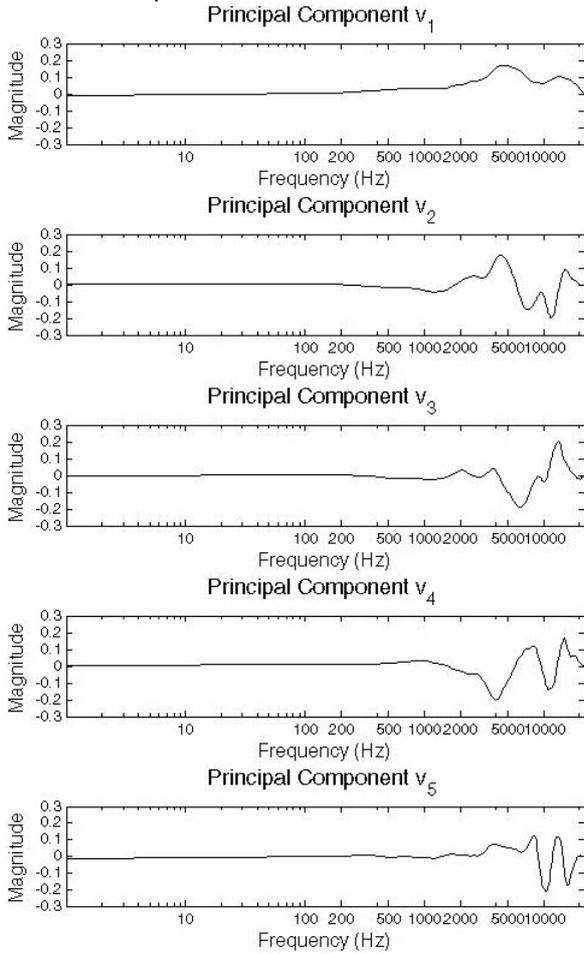

Fig. 5. The first five basis functions or PCs extracted from PCA of 3700 DTFs from both ears of 37 subjects on horizontal plane.

### 3.2 Weights of Basis Functions

Based on PCA, assumed that DTFs can be represented by a relatively small number of basic spectral shapes of PCs, it seems reasonable to expect that the amount each basic shape contributes to the DTF at a given source position would related, in a simple way, to source azimuth and elevation. In the case of source position on horizontal plane, this amount or weight is related to azimuth only.

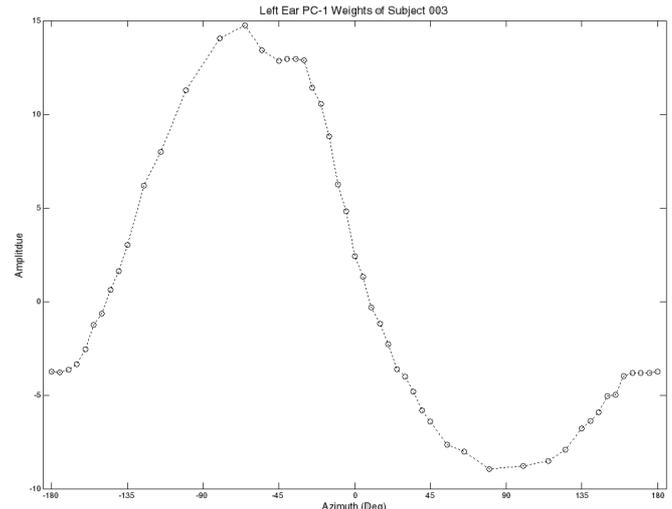

Fig. 6. Left ear PC-1 weights for DTFs of Subject 003 on horizontal plane.

Fig. 6 shows left ear PC-1 weights for DTFs of Subject 003, which were plotted as function of source azimuth on the front horizontal plane. It is shown, that there is a tendency for the weights to decrease in magnitude as the source moves from the median plane (azimuth 0º). Ipsilateral sources, sources with negative azimuths, have positive weights with exception the sources with azimuth -180º to -150º have negative weights. On the other side, contralateral sources have negative weights. The magnitudes of weights for ipsilateral sources are much larger than those for contralateral sources. This distribution of PC-1 weights is similar across the 37 subjects, which has low intersubject variability. As seen in Fig. 5, that the first basis function has almost flat magnitude through all frequencies, so it can be said that PC-1 weights are functioning as the amplification in HRTF modeling.

The remaining nine PC weights have larger variability in the ipsilateral side and, in the contralateral side, beginning at about azimuth 0º, the weights have almost constant values near zero. Higher order of PC has corresponding flatter weights pattern for sources on the horizontal plane. It is observed, that the patterns of PC weights are roughly similar across subjects and ears.

### 3.3 Performance of Proposed Individualization Method

The performances of the estimated magnitude HRTFs on the horizontal plane, obtained either from PCA or individualization, were evaluated by the comparison of mean-square error of the differences between the estimated magnitude HRTFs, and the original magnitude HRTFs, calculated from database, to the mean-square error of the original magnitude HRTFs in percentage, which is defined by,

$$e_j(\theta) = 100\ \% \times \| \mathbf{h}_j(\theta) - \hat{\mathbf{h}}_j(\theta) \|^2 / \| \mathbf{h}_j(\theta) \|^2 \qquad (16)$$

where $\mathbf{h}_j(\theta)$ is the j-th original magnitude HRTF with azimuth $\theta$ on horizontal plane, $\hat{\mathbf{h}}_j(\theta)$ is the corresponding estimated magnitude HRTF of $\mathbf{h}_j(\theta)$. If the error is larger, the performance of the estimated magnitude HRTF is worse, where better localization results will be achieved with small error, $e_j(\Omega)$.

Before individualizing magnitude HRTFs using MLR, mean error from PCA modeling of magnitude HRTFs was calculated across all data in the database. At first, PCA modeling was performed for all data from all source directions of 45 subjects. This experiment resulted in mean error of 3.31% across all directions and subjects, but mean error across directions on horizontal plane was 3.65%. Second, modeling was performed using data at all directions of only 37 subjects, which resulted in mean error of 3.32% and mean error on horizontal plane was 3.68%. From these two experiments, it can be said that the corresponding mean errors were the same. Third, data of both ears of 45 subjects at directions only on horizontal plane were used and mean error of 3.67% was obtained. At last, PCA modeling was performed using data of both ears of only 37 subjects at directions only on horizontal plane. This experiment resulted in mean error of 3.68%. Again we obtained the same mean errors from the last two experiments. It is summarized that using data of 45 subjects or 37 subjects, yielded the same mean errors across related directions. Mean errors on horizontal plane were the same either data from all directions used or only from directions on horizontal plane. These mean errors are less than half of the related mean errors obtained from our previous work on PCA modeling of minimum phase HRIRs[14].

In individualizing magnitude HRTFs, we used only the data of both ears of 37 subjects at directions on horizontal plane, which meant that we used the results of fourth experiment mentioned above for individualization. We obtained here significantly small mean error of PCA models of magnitude HRTFs, i.e. 3.68% compared to 8.32% as in [14].

In turn, we individualized the PCA model of magnitude HRTFs using MLR with eight chosen measurements. The mean error of a subject was different from that of another subject in the database and also noted that a good performance of the individualized left-ear magnitude HRTFs of a subject was not always followed by the same performance of the right-ear ones. The overall mean error was only 12.17%, which was much better than 22.50% as in [14]. Fig. 7 shows the left- and right-ear errors as a function of azimuths on the front horizontal plane of subject 003, in the top panel, and of subject 163, in the bottom panel, after individualizing magnitude HRTFs. The mean errors for both ears of subject 163 are worse than those for both ears of subject 003, i.e. 12.92% and 21.20% repectively for left ear and right ear of subject 163, but only 8.27% and 5.18% repectively for left ear and right ear of subject 003. The left-ear errors of subject 003 on the front horizontal plane are about 10%, except for azimuth 65°, where positive azimuth is due to source direction in the right side. However, the right-ear errors of subject 003 are mostly very good about 5% across azimuths. The left-ear errors of subject 163 seem to be much better than its right-ear errors.

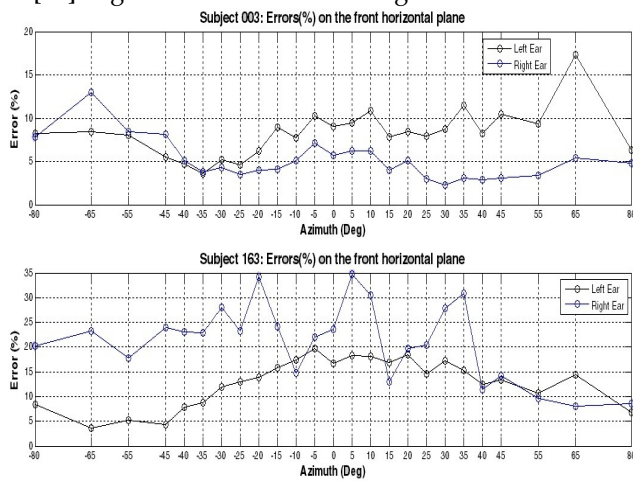

Fig. 7. Left- and Right-Ear Errors of Subject 003 and Subject 163 on the Front Horizontal Plane After Individualization

Under the assumption stated below, if the spectral distortion (SD) score defined by Hu et al.[12], was applied to determine the performance of the individualized magnitude HRTFs, our SD scores of subject 003 and subject 163 on the front horizontal plane are no larger than 1 dB, which is much better than that in [12]. Using logarithm properties, i.e. $20.\log(|a|/|b|) = (20.\log|a| - 20.\log|b|)$, we assumed that the difference of log-magnitudes $(20.\log|a| - 20.\log|b|)$ in SD score might be replaced by the difference of magnitudes $(\log|a| - \log|b|)$ in our case because we resulted in individualized magnitudes of HRTFs and Hu et el. resulted in individualized log-magnitudes of HRTF from PCA.

There were overall additional errors introduced by the proposed individualization method. These additional errors were introduced by the MLR. The unsystematic behavior of weights of PCs across subjects and across directions had caused MLR quite difficult to estimate adequately accurate regression coefficients. Besides, we performed here linear regression of anthropometric measurements to estimate the weights of PCs. Higher order regression might provide better estimates of these weights.

The individualized magnitude HRTFs of subject 003 could well approximate the corresponding original magnitude HRTFs particularly at frequencies below about 8 kHz. Fig. 8 shows the individualized and original magni-tude HRTFs for both the left and right ear in the extreme directions on the front horizontal plane. The top, middle, and bottom panel corresponds to azimuth -80°, 0°, and 80° respectively.

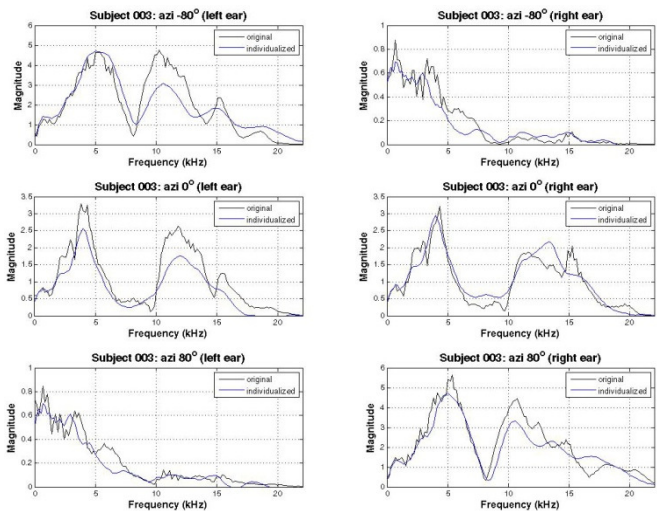

Fig. 8. Magnitude Responses of the Original and Individualized HRTFs of Subject 003 on the Front Horizontal Plane.

Informal listening tests done by five subjects had shown a good and natural perceived moving sound around the horizontal plane by all subjects when the subjects' individualized reconstructed HRIRs, due to the sound source directions, were implemented in the headphone simulation.

## 4 CONCLUSION

In this paper, a simple and efficient individualization method of magnitude HRTFs for sources on horizontal plane, based on principal components analysis and multiple linear regression, was proposed. The proposed method shows better performance in the objective simulation experiments than that of similar research and was superior compared to our previous work. The additional errors introduced by MLR to PCA model might be lowered by applying higher order regression or other algorithm for MLR than the least square.

### ACKNOWLEDGMENT

The authors wish to thank all staffs of CIPIC Interface Laboratory of California University at Davis, USA for providing the CIPIC HRTF Database. This work was supported in part by a grant from Electrical Engineering Department, Industrial Technology Faculty, Trisakti University, Jakarta, Indonesia.

**Hugeng** received the bachelor and master degrees in electrical engineering from Trisakti University, Indonesia, in 1995 and 1998 respectively. He worked as research assistant at Institute of Informatic VI, RWTH Aachen, Germany, and at Institute for Digital Signal Processing, TU Kaiserslautern, Germany. He is recently a lecturer at Electrical Engineering Department, Trisakti University, Indonesia. He is now pursuing towards the Ph.D degree in Electrical Engineering Department, University of Indonesia, Indonesia.

**Wahidin Wahab** received the B.Sc. degree in electrical engineering from University of Indonesia in 1978, and M.Sc and Ph.D degrees from University of Manchester Institute of Science and Technology, England in 1983 and 1985, respectively. He is a lecturer in Electrical Engineering Department, University of Indonesia. His research interests are robotics, intelligent system, and signal processing. Dr. W. Wahab is a senior member of IEEE.

**Dadang Gunawan** received the B.Sc. degree in electrical engineering from University of Indonesia in 1983, and M.Eng and Ph.D degrees from Keio University, Japan, and Tasmania University, Australia in 1989 and 1995, respectively. He is the Head of Telecommunication Laboratory and of the Wireless and Signal Processing Research Group of the Electrical Engineering Department, University of Indonesia. His research interests are wireless communication and signal processing. Prof. Dr. D. Gunawan is a senior member of IEEE and IEEE Signal Processing Society.